\documentclass[prb,twocolumn,showpacs,superscriptaddress,floatfix, nofootinbib]{revtex4-2}
\usepackage{amsmath,amssymb,amsfonts,mathrsfs, amsbsy}
\usepackage{graphicx}
\usepackage{lipsum, babel}
\usepackage[bookmarks=true, colorlinks=true, linkcolor=blue, urlcolor=blue, citecolor=blue, bookmarks=true, hyperindex=true]{hyperref}
\usepackage{comment}

\newcommand{\be}{\begin{equation}}
\newcommand{\ee}{\end{equation}}

\newcommand{\beq}{\begin{eqnarray}}
\newcommand{\eeq}{\end{eqnarray}}

\def\H1{\widehat{H}_1}

\begin{document}

\title{Enhanced Drude weight in a 1D system of fermions with pair hopping events}

\author{M.\,S.~Bahovadinov}
\affiliation{ Physics Department, National Research University Higher School of Economics, Moscow 101000, Russia}
\affiliation{Russian Quantum Center, Skolkovo, Moscow 143025, Russia}
 \author{R.\,O.~Sharipov}
 \affiliation{ Physics Department, Faculty of Mathematics and Physics, University of Ljubljana,
Ljubljana, Slovenia}
 \affiliation{Russian Quantum Center, Skolkovo, Moscow 143025, Russia}
   \author{B.\,L.~Altshuler}
   \affiliation{Russian Quantum Center, Skolkovo, Moscow 143025, Russia}
  \author{G.\,V.~Shlyapnikov}
\affiliation{Russian Quantum Center, Skolkovo, Moscow 143025, Russia}
\affiliation{Moscow Institute of Physics and Technology, Dolgoprudny, Moscow Region, 141701, Russia}
\affiliation{Université Paris-Saclay, CNRS, LPTMS, 91405 Orsay, France}
\affiliation{Van der Waals–Zeeman Institute, Institute of Physics, University of Amsterdam, Science Park 904, 1098 XH Amsterdam, The Netherlands}
\date{\today}

\begin{abstract}
In one dimension density-density interactions of particles reduce their mobility and hence the Drude weight, which controls the divergence of the optical conductivity at zero frequency, decreases. We study effects of pair hopping events on this result in a 1D system of spinless fermions. The considered model consists of the usual single-particle hopping and pair hopping terms. In the absence of the density-density interactions, we first show that a variation of the pair hopping amplitude results in a monotonic change of the Drude weight. We next demonstrate that weak nearest-neighbor density-density interactions increase the Drude weight, whereas in the regime of strong interactions the Drude weight decreases as expected. Our numerical findings are supported by bosonization results.
 \end{abstract}

\maketitle

\section{Introduction}  
One of the distinguished differences of quantum-mechanical systems from their classical counterparts is the existance of the Aharonov-Bohm effect. A charged particle acquires a phase when its trajectory encloses a finite magnetic flux. The enclosed magnetic flux $\Phi$ is proportional to the vector potential along the looped trajectory and it perturbs energy levels of the quantum-mechanical system. The manifestation of this effect occurs in quasi-1D mesoscopic conducting rings of length $L$, where a dissipationless persistent current $J\propto\frac{{\cal{D}}\Phi}{L}$ is induced in response to a magnetic flux $\Phi$ threading the ring~\cite{Buttiker,Cheung,Levy,Shanks,Schmid,Kulik,Viefers,Bouchiat,Imry}. The prefactor $\cal D$ is the Drude weight of the system, which also determines the zero-frequency divergence of the conductivity~\cite{Shastry,Fye,Scalpino,GiamShast},
\be
\label{Conductivity_expr}
\Re\left[ \sigma\left(\omega\right) \right]=\pi {\cal D}\delta(\omega) + \sigma_{reg}(\omega).
\ee
This quantity was first established by Kohn~\cite{Kohn} to quantify conducting properties of strongly correlated many-body systems. For conducting systems, which support ballistic transport of constituent carriers, one has ${\cal D}>0$, whereas for insulating systems one has ${\cal D}=0$. The Drude weight is formally introduced as a response of the ground state energy to an infinitesimal magnetic flux: 
\be
\label{Fdiff}
{\cal D}=L\frac{d^2E_0}{d\Phi^2}\rvert_{\Phi \rightarrow0}.
\ee
In quasi-1D systems the Drude weight ${\cal{D}}$ and the superfluid density (spin or charge stiffness) $\rho_S$ are equal to each other at zero temperature and can be interpreted as equivalent quantities. Quantitative estimation of the Drude weight is important for the interpretation of experiments addressing persistent currents in mesoscopic metallic rings~\cite{Millis,Schmid,Imry,Bouchiat} and persistent flows in setups of ultracold atoms in optical traps~\cite{Sauer,Gupta,Ryu,Lesanovski,Eckel,Pichler, Amico,Rossini,Gallemi,Mancini,Genkina}. Therefore, it is important to classify effects of strong correlations on the Drude weight in quantum many-body systems.     
The generally accepted scenario is that repulsive interparticle interactions always reduce particle mobility of a generic quantum many-body system and, hence, they lead to a reduced Drude weight~\cite{Meden, Dias, Bouzerar, Berkovitz}. However, recent studies~\cite{Bischoff,Haller} of quasi-1D systems in a ladder geometry have shown unusual effects of interparticle interactions on the Drude weight. In ladder systems with a transverse magnetic flux, the Drude weight increases linearly with the amplitude of repulsive interactions, whereas for attractive interactions the Drude weight decreases. Such a counterintuitive result is guaranteed by the caused bias between back- and forward-scattering processes. This bias results from an interplay between interparticle interactions  and the transverse magnetic flux. 

 One of the possibilities to enhance particle mobility in lattice systems is to introduce pair hopping of particles. Recently, a model with this feature was studied in 1D by J. Ruhman and E. Altman~\cite{RuhmanAltman}. Although the Ruhman-Altman model is rather abstract, it got sufficient attention and the phase diagram of this model was recently studied numerically by means of the Density Matrix Renormalization Group (DMRG) method ~\cite{Mazza,Mazza2}. However, effects of pair hopping events on transport properties remain an open subject. 

In this work we present another interesting model with an imposed pair hopping, where the Drude weight also monotonically changes when varying the system parameters. Namely, we consider 1D spinless fermions with single-particle and pair hoppings. Using numerical DMRG~\cite{White1,White2,Schollwock,Rizzi} and bosonization of the model, we show that the Drude weight monotonically varies with the pair-hopping amplitude $J_2$. For positive $J_2$ it decreases with $J_2$, whereas for negative $J_2$ it increases with $|J_2|$(linearly for $|J_2 |\ll 1$). We also show that an additional attractive (repulsive) nearest-neighbor density-density interaction term further enhances the Drude weight in the regime where the pair-hopping events are energetically favoured (disfavoured).
This is in sharp contrast to the case of $J_2=0$, where no renormalization of the Drude weight is achieved at the leading order in a system with such interactions and higher corrections usually result in a decrease of the Drude weight. 
The model of our study can be mapped onto the $J_1-J_2$ XY model via the Jordan-Wigner transformation~\cite{JWT} and can be experimentally realized in several systems (using 3D transmon qubits on saphire and also using ultracold bosonic atoms in optical lattices), as suggested by recent proposals~\cite{Proposal1,Proposal2}.

\begin{figure}[t]
\includegraphics[width=\columnwidth]{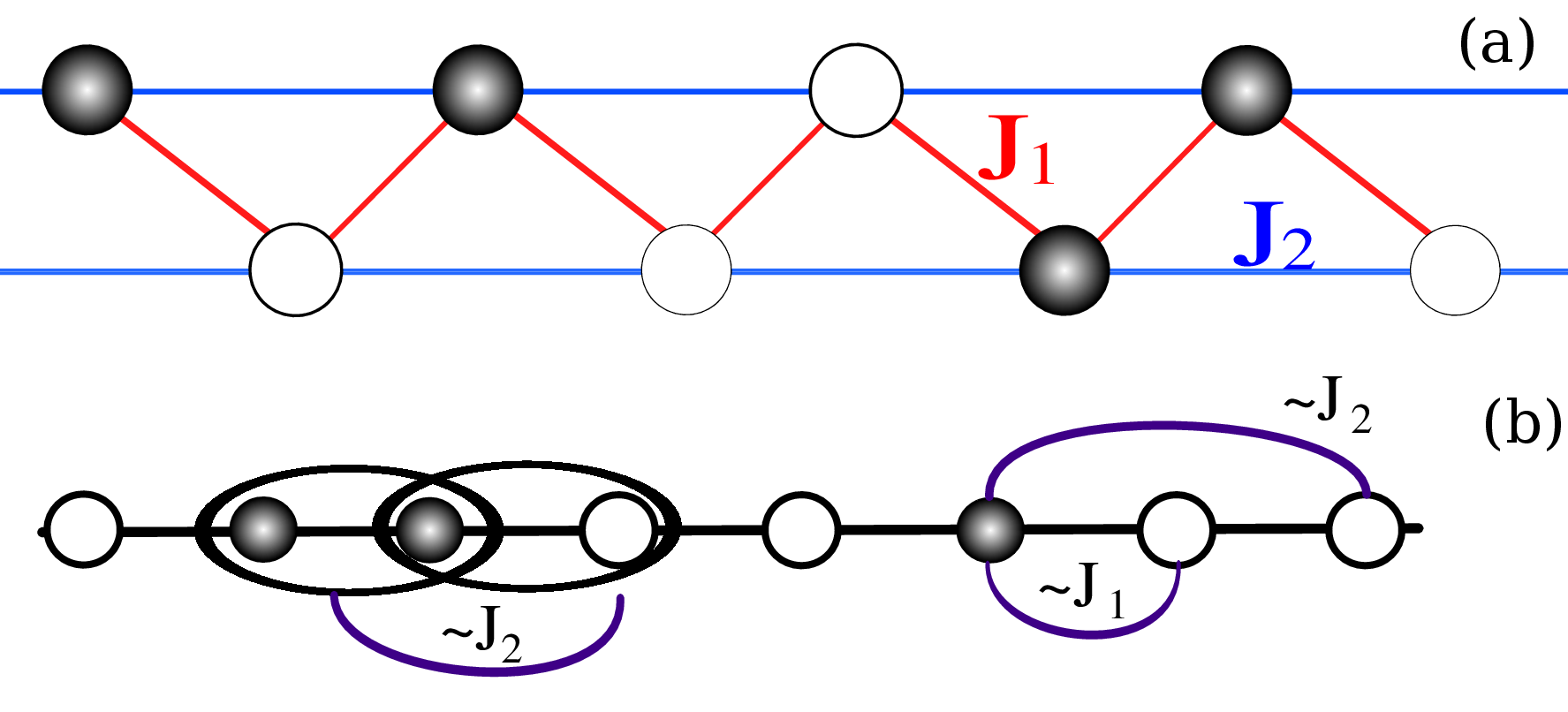}
\caption{(a) Schematic representation of the fermionic model (\ref{H}). For $J_2$/$|J_1|<0$ on top of single particle hoppings pair hopping events are also favoured, as sketched in (b).}
\label{fig:r}
\end{figure}

The paper is organized as follows. In Sec.~\ref{S_model_spin} we present the fermionic model and discuss its main properties. To highlight the key results of our work, in Sec.~\ref{Sec3} we emphasize why repulsive density-density interactions do not result in a strong (linear in the interaction strength) renormalization of the Drude weight in 1D systems. In Sec.~\ref{Sec4} we present the low-energy Tomonaga-Luttinger liquid (TLL) theory of the model, obtained by both constructive  and field-theoretical bosonization and show strong renormalization of the Drude weight within the TLL theory. To validate the bosonization results we present our numerical findings of the excitation velocity and the TLL parameter $K$ in Sec.~\ref{Sec5}. In Sec.~\ref{Sec6} we demonstrate our numerical findings for the Drude weight in all considered regimes. Sec.~\ref{Conc} is devoted to our concluding remarks.

\section{Model and symmetries} 
\label{S_model_spin}

We consider a 1D system of spinless fermions with single-particle and pair hopping terms, defined on the zig-zag ladder of (even) $L$ sites with periodic boundary condition (see Fig.~(\ref{fig:r})). The Hamiltonian

\be \label{H}
	H =  H_1+H_2 
\ee
contains the single-particle hopping term $H_1$ and pair hopping term $H_2$:
\be \label{H_1}
	H_1 =  \sum_{\beta=1,2} \sum_{j=1}^{L} \frac{J_\beta(-1)^{\beta+1}}{2} \left(c^\dagger_i c_{i+\beta} + h.c. \right), 
\ee
\be \label{H_2}
H_2=J_2\sum_i^{L} \left( c^\dagger_{i}c_{i+1} c^\dagger_{i+1}c_{i+2} +h.c. \right),
\ee
where we impose $c_{L+1}=c_1$. 
For convenience we also introduce a parameter $\kappa=\frac{J_2}{|J_1|}$ and consider half-filled ladder.   

The Hamiltonian (\ref{H}) is translationally invariant and conserves the particle number (U(1) symmetry) and parity symmetries. If $\kappa=0$ or $\kappa=\infty$, the system is integrable, since one can map the model exactly onto the model of free fermions in 1D.  
For $\kappa \ll 1 $ the clean model is quasi-integrable, possessing
quasi-conserved charges, as was shown in the recent works~\cite{Kurlov2021,Bahovadinov2022,Kurlov2023,Motrunich2023}. We consider only $-1< \kappa \lesssim 0.33$, where the TLL theory serves as a valid framework of $T \rightarrow 0$ physics. We emphasize that the Hamiltonian (\ref{H}) can be mapped exactly via Jordan-Wigner transformation~\cite{JWT} onto the XY model ($s=1/2$) in the zig-zag ladder~\cite{Bahovadinov2022}:
\be \label{H_spin}
 H =  \sum_{\beta=1,2}J_\beta \sum_{i=1}^{L} \left[ S^x_i S^x_{i+\beta}+S^y_i S^y_{i+\beta} \right].
 \ee
 Recently, it was shown in Ref.~\cite{Bahovadinov2022} that the current model with on-site disorder exhibits MBL transition provided by the pair-hopping term (\ref{H_2}) in the fermionic formulation. A strong pair hopping also guarantees a superfluid (SF)-Bose glass transition at finite disorder, as it was demonstrated in Ref.~\cite{Bahovadinov2023}. 

At $T=0$ one expects that for $\kappa<0$ the SF phase is conserved with the modified TLL parameter $K>1$. Indeed, pair-hopping of fermions amplifies SF correlations, resulting in a slower algebraic decay of these correlations. On the contrary, for $\kappa>0$, one has dominating charge-density wave type correlations, with $K<1$. To highlight this, one can rewrite the term from (\ref{H_2}) as a correlated hopping term $-\kappa(c^\dagger_in_{i+1}c_{i+2} +h.c.)$. Hopping of fermions along a given leg pins fermionic density on the other leg. If $\kappa>0$ and is sufficiently large, then such pinning can cause spontaneous dimerization onto the $2k_F$ bond-order density wave~\cite{Haldane1982}. This phase transition was previously shown~\cite{Haldane1982,Okamoto1993,Hirata1999, Lecheminant2001,Sota2010,Mishra2013, Bahovadinov2023} to occur in the vicinity of $\kappa_c \approx 0.33$.  
\section{Weak renormalization of the Drude weight in 1D}
\label{Sec3}
In one dimension interparticle interactions do not affect the Drude weight in Galilean invariant systems due to the decoupling of the center-of-mass motion from internal degrees of freedom~\cite{GiamarchiBook,Haller,GogolinBook,Maslov}. The presence of a lattice changes this result only beyond the leading order.
To show this, let us consider non-interacting fermionic theories in 1D, which are successfully described within the bosonization formalism (we set $\hbar=e=1$): 
\be
\label{H0}
H_0=\frac{v_F}{2}\int dx \left[ \left(\partial_x\phi\right)^2+\left(\partial_x\theta\right)^2 \right],
\ee
where $v_F$ is the Fermi velocity and the conjugated bosonic fields $\phi(x)$ and $\theta(x)$ satisfy commutation relations
\be
\left[\phi(x),\partial_{x^\prime}\theta(x^\prime) \right]=i\delta(x-x^\prime).
\ee
In the presence of short-range density-density interactions of the form $H_{int}=\sum_{i,j} V_{i,j}\hat{n}_i\hat{n}_j$, Eq.~(\ref{H0}) takes the form 
\be
H=\frac{v_s}{2}\int dx \left[ \frac{1}{K}\left(\partial_x\phi\right)^2+K\left(\partial_x\theta\right)^2 \right]+{\cal{V}}(\phi),
\ee
where $v_s$ is the new excitation (sound) velocity and $K$ is the TLL parameter. The parameter $K$ is smaller than unity for generic $V_{i,j}>0$, whereas $K>1$ for attractive interactions. Importantly, the term ${\cal V}(\phi)$, which includes all non-quadratic terms, only depends on $\phi$ and does not depend on $\theta$. Hence, $\left[\hat{n},H_{int}\right]=0$, since the electronic density operator is also only a function of $\phi$. From the continuity equation $\partial_t\hat{n}=i\left[H,\hat{n}\right]=-\partial_xj$ one obtains the expression for the current 
\be
j=-\frac{v_F}{\sqrt{\pi}}\partial_x\theta.
\ee
The current is not effected by the density-density interactions and the Drude weight remains equal to the non-interacting Fermi velocity $v_F/\pi$:
\be
\label{D_0}
{\cal D}=\frac{v_sK}{\pi}=\frac{v_F}{\pi}={\cal D}_0.
\ee
The terms in ${\cal{V}}(\phi)$ are usually irrelevant in the renormalization group sense and lead to the corrected parameters $v_s^*$ and $K^*$. The Drude weight also gets renormalized, 
\be
{\cal D}=\frac{v_s^*K^*}{\pi}<{\cal{D}}_0.
\ee
However, it is important to emphasize that such renormalization is usually weak, at most of the second order in the perturbative expansion, and the Drude weight decreases. 

Alternative way of explanation of the above fact can be done in terms of $g$-ology approach: $H_{int}$ causes backscaterring process with the amplitude $g_2$ and the forward scattering process with the amplitude $g_4$. The Drude weight in terms of these amplitudes takes the following form~\cite{GiamarchiBook}:
\be
{\cal{D}}={\cal D}_0+\frac{g_4-g_2}{2\pi^2}.
\ee
 For the same reasons leading to Eq.~(\ref{D_0}), for the density-density interactions one always finds $g_2=g_4$ and no renormalization of the Drude weight occurs at the leading order. 

The typical example of the described scenario occurs in the 1D spin-1/2 XXZ model with the  Ising interactions $H_{int}=\sum_i J^z S^z_iS^z_{i+1}$. This model can be mapped onto the model of 1D spinless fermions with the nearest-neighbor hopping and the density-density interactions
\be
H_{int}=\sum_iJ^z(\hat{n}_i-1/2)(\hat{n}_{i+1}-1/2).
\ee
At $|J^z|\ll1$, the standard bosonization procedure leads to $g_2=g_4=4J^z$ and the parameters
\be
K=v_F/v_s=\left(1-\frac{4J^z}{\pi}\right)^{-1/2}.
\ee
Thus, the Drude weight is equal to ${\cal{D}}_0$ and it is not renormalized at the leading order. However, irrelevant terms encapsulated in ${\cal{V}}({\phi})$ result in the decrease of the Drude weight for both repulsive ($J^z>0$) and attractive ($J^z<0$) interactions. This can be shown using the exact form of the parameters $v^e_s$ and $K^e$ known from the Bethe Ansatz~\cite{YangYang66}, 
\be
K^e=\frac{\pi}{2}\frac{1}{\pi-\arccos{J^z}}
\ee
and 
\be
v_s=\frac{\pi}{2}\frac{\sqrt{1-J_z^2}}{ \arccos{J^z}}.
\ee
Thus, the exact Drude weight is ${\cal D}=\frac{v_s^eK^e}{\pi}<{\cal D}_0$ for $|J^z|<1$ (see Fig.~\ref{fig:Drude1}(c) for the plot of the Drude weight as a function of $J^z$).
What is important for our next discussions is that the renormalization of the Drude weight for the density-density interactions in 1D is not exhibited within the framework of the TLL theory, but only irrelevant terms lead to weak corrections. As we show in the next sections, in the model of our study this scenario is violated and large corrections to the Drude weight due to the pair hopping events already manifest themselves on the TLL theory level. We also demonstrate numerically that in the presence of pair hopping density-density interactions (which usually result in the decrease of the Drude weight) enhance the Drude weight. We provide qualitative arguments for these results.

\section{Bosonization procedure}
\label{Sec4}
{\it{Constructive bosonization}} - We first follow a constructive bosonization procedure to achieve an effective low-energy theory of the model. For this  the Hamiltonian (\ref{H}) is rewritten in the $k$-space:
\be\label{Hk}
\Tilde{H}=\sum_{k\in BZ} \epsilon_k c^\dagger_k c_k + \frac{J_2}{L}\sum_{k_1,k_2,q} \cos(2k_1+q)c^\dagger_{k_1+q}c_{k_1}c^\dagger_{k_2-q}c_{k_{2}},
\ee
with the single-particle dispersion relation:
\be
\epsilon_k=-\sum_{\beta=1,2} J_\beta\cos(\beta k).
\ee
In the weak-coupling regime, $|\kappa|\ll1$,  one starts with a linearized spectrum of the free fermionic term (\ref{H_1}) with the corresponding left (L) and right (R) moving branches. The first term of (\ref{Hk}) can be rewritten as, 
\be \label{H0RHO}
\Tilde{H_0}=\frac{\pi v_F}{L} \sum_{q,\tau} \hat{\rho_{\tau}}(q)\hat{\rho_{\tau}}(-q),  
\ee
with $\tau \in [ L(-1), R(+1) ]$, where the Fermi velocity is $v_F=\frac{\partial\epsilon_k}{\partial k}|_{k=k_F}$ and the density plasmons for a given branch $\tau\in L,R$ are defined as,
\begin{equation} 
\hat{\rho}_\tau(q)=\sum_k c^\dagger_{\tau,k+q} c_{\tau, k}.
\end{equation}
Canonical fermionic operators $c^{(\dagger)}_{k,\tau}$ correspond to the $\tau$ branch. The second term of Eq.~(\ref{Hk}) can not be directly expressed in terms of these plasmons due to the $k$-dependence of the amplitude $V(k,q)=\cos(2k+q)$. However, for $|\kappa| \ll 1$  one can assume that  $V(k,q)\approx V(k_F,q)$, since the momentum of excitations $q$ is close to zero for the forward scattering, and $q\sim 2k_F$ for the back-scattering processes. One is left with the $k_F$ dependence of the scattering amplitudes $V(q\sim 0)=\cos(2k_F)$ and $V(q\sim 2k_F)=\cos(4k_F)$. This is expected, since if the density of particles (holes) exceeds half-filling, pair-hopping events are less probable and the effects of the corresponding term are weak, i.e the largest contribution is expected at half-filling. Within this approximation, one can rewrite the second term of Eq.~(\ref{Hk}) in terms of the plasmonic excitations and fully bosonize the fermionic theory, since $[\hat{\rho}_\tau(-q),\hat{\rho}_{\tau^\prime}(q^\prime)]=\frac{Lq\tau}{2\pi}\delta_{\tau,\tau^\prime}\delta_{q,q^\prime}$. 
The second term of the Hamiltonian in terms of density plasmons can be expressed as
\be
H_{int}=H_{g_4}+H_{g_2},
\ee
where
 \be
H_{g_4}=\frac{1}{2L}\sum_{q\sim 0} g_4 \left( \hat{\rho}_{R,-q}\hat{\rho}_{R,q}+\hat{\rho}_{L,-q}\hat{\rho}_{L,q} \right),
 \ee
and 
\be
H_{g_2}=\frac{1}{2L}\sum_{q\sim 0} g_2 \left( \hat{\rho}_{R,-q}\hat{\rho}_{L,q}+\hat{\rho}_{L,-q}\hat{\rho}_{R,q} \right). 
\ee
 The amplitudes $g_4$ and $g_2$ are given by 
\be
g_4=8\kappa \cos(2k_F),
\ee
 and
 \be
 g_2=4\kappa (1-\cos(2k_F)).
 \ee

One then follows the standard bosonization scheme~\cite{GogolinBook, GiamarchiBook, Maslov} by introducing the conjugated bosonic fields,
\be
\phi(x)=i\sum_{q\neq0} \frac{sgn(x)}{\sqrt{2|q|L}}(b^\dagger_qe^{-iqx}-b_qe^{iqx}),
\ee
and
\be
\theta(x)=-i\sum_{q\neq0} \frac{1}{\sqrt{2|q|L}}(b^\dagger_qe^{-iqx}-b_qe^{iqx}),
\ee
with $[\phi(x),\partial_{x^\prime}\theta(x^\prime)]=i\delta(x-x^\prime)$. As a result, the (1+1) dimensional Sine-Gordon model is obtained,
\be
\label{SG}
H=\frac{v_s}{2}\int \left( \frac{1}{K}(\partial_x\phi)^2+K(\partial_x\theta)^2 \right) + g \cos(\beta_s\phi),
\ee
where $\beta_s=\sqrt{16\pi}$, and $v_s$ is the excitation (sound) velocity. The cosine term originates from the $4k_F$ umklapp scattering, since we consider the half-filled sector of the Hilbert space. General expressions for the TLL parameter $K(\kappa,k_F)$ and the excitation velocity $v_s(\kappa,k_F)$ have the following forms:
\be
K(\kappa,k_F)=\sqrt{\frac{2\pi+4\kappa\left(3\cos(2k_F)-1\right)}{2\pi+4\kappa\left(\cos(2k_F)+1\right)}},
\ee
and 

\be
v_s(\kappa,k_F)=v_F\sqrt{\left(1+\frac{4\kappa\cos(2k_F)}{\pi}\right)^2-\left(\frac{4\kappa \sin(k_F)}{\pi}\right)^2  }.
\ee 
At half-filling ($k_F=\frac{\pi}{2}$) the expression for $K$ and $v_s$ transforms into
\be
\label{K_expression}
\frac{v_s}{v_F}=K=\sqrt{1-\frac{8\kappa}{\pi}}.
\ee
 
For $K<1/2$ the cosine term in Eq.(\ref{SG}) becomes relevant in the RG sense and opens a gap in the spectrum via the BKT transition. From Eq.(\ref{K_expression}) we find the critical value  $\kappa_c=\frac{3\pi}{32} \approx 0.295.$ Recently, using DMRG the critical  $\kappa_c=0.3256(2)$ was obtained~\cite{Bahovadinov2023}, which is in good agreement with the bosonization result.

The expression for the Drude weight can be obtained directly using Eq.(\ref{K_expression}):
\be
\label{Drude_weight_expr}
{\cal{D}}=\frac{v_sK}{\pi}={\cal{D}}_0+\frac{g_4-g_2}{2\pi^2}=v_F\left(\frac{1}{\pi}-\frac{8\kappa}{\pi^2}\right).
\ee

Such a linear renormalization of the Drude weight is one of the main results of this work. It is a consequence of the broken Galilean invariance of the model. On the constructive bosonization level, the latter is manifested as $g_4\neq g_2$. 

{\it{Field-theoretical bosonization}} - Alternative way of explaining the previous facts is to use the field theoretical bosonization (see Appendices A and B for details). In the field-theoretical bosonization, one starts with the fermionic field expression for the hopping term:
\begin{equation}
\begin{split}
    H_{int}= & -J_2
\sum_n c_{n+1}^{\dagger} :c^{\dagger}_n c_n: c_{n-1}+h.c. \rightarrow \\ 
&-J_2 a \int dx~ \psi^{\dagger}(x+a):\psi^{\dagger}(x) \psi(x):\psi(x-a)+h.c.,
\end{split}
\end{equation}
where $a$ is the lattice constant and $::$ denotes  normal ordering of the fields.
Decomposing the fermionic fields into the right and left moving fields, using bosonization identity and neglecting oscillating terms, one can obtain:

\begin{equation}
\begin{split}
        H_{int}=&-\frac{ 4~a J_2}{\pi} \int dx \left(\partial_x \theta \right)^2\\ & -\int dx \frac{ a J_2 }{(\pi \alpha)^2} \cos{\left(\sqrt{16 \pi} \phi \right). }
        \end{split}
\end{equation}
While the second (Umklapp) term does depend on $\phi(x)$ field, one clearly notices that the first term depends on $\theta(x)$ field also. This breaks the commutation relation $\left[\hat{n},H_{int} \right]\neq 0$, and hence the Drude weight also gets strongly renormalized. This fact accomplishes the previous conclusion of the strong Drude renormalization obtained within the constructive bosonization procedure. In the following sections, we check these findings numerically.  
 \begin{figure}[t]
\includegraphics[width=\columnwidth]{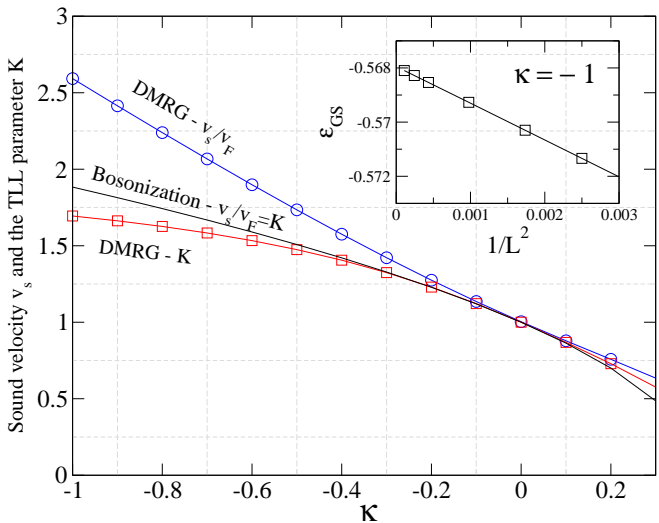}
\caption{ Numerical DMRG results for the sound velocity $v_s$ and the TLL parameter $K(L=128)$ versus $\kappa$. The inset shows the results of the fitting procedure of Eq.~(\ref{Scaling}) at $\kappa=-1$.  }
\label{fig:VandK}
\end{figure}

\section{Numerical results: critical properties}
\label{Sec5}
It is important to check the validity of Eq.~(\ref{K_expression}) by means of numerical DMRG method.
 In this section we present details of our numerical calculations and show our results for the sound velocity and the TLL parameter $K$. For numerical convenience we considered the model in its spin-1/2 representation (\ref{H_spin}).
    We used variational single-site DMRG algorithm to obtain accurate matrix product state representation of the ground state in the half-filled sector of the Hilbert space from a given product state. We performed a large number of sweeps and gradually increased bond-dimensions up to $\chi \approx 1000$ for the largest considered system sizes. These steps guarantee the convergence to the ground state during the RG procedure. Largest truncation errors in the last sweepings were of the order of $\epsilon\sim 10^{-10}$. 
  
{\it{Sound velocity $v_s$.}} Following the standard level spectroscopy methods of Conformal Field Theories~\cite{Xavier,Taddia}, the excitation velocity $v_s$ of the single component TLL can be extracted from the finite-size scaling of the ground state energy density (using periodic boundary conditions):
\be
\epsilon_{GS}(L)=\epsilon_{0}+\frac{\pi v_s c}{6L^2}+...,
\ee
where $\epsilon_0$ is the thermodynamic value of the ground state energy density, and $c$ is the central charge ($c=1$ within the TLL phase of our model, see Ref.~\cite{Bahovadinov2023} for details). The sufficient number of system sizes considered in
our calculations allows to safely perform the following fit: 
\be
\label{Scaling}
\epsilon_{GS}(L) = a_0 + \frac{a_1}{L^2}+ \frac{a_2}{L^{a_3}}.
\ee
An example of the fitting result at $\kappa=-1$ is shown in the inset of Fig.~\ref{fig:VandK}. The extracted values of the excitation velocities as  functions of $\kappa$ are also presented in Fig.~\ref{fig:VandK} (blue circles). The fitting procedure correctly results in $v_s=v_F$ at $\kappa=0$. For $\kappa<0$ the energetically favoured pair-hopping events lead to the increased values of $v_s$, reaching $v_s/v_F\approx2.5$ at $\kappa=-1$. At positive values of $\kappa$ one has dominant density-wave correlations in the bulk of the system and one has $v_s/v_F<1$. It is important to emphasize that although there are large discrepancies between the numerical DMRG and bosonization result (solid line) at large $\kappa$, for $|\kappa|< 0.2 $ the bosonization result nicely overlaps with the numerical data. This is expected, since the bosonization of the model was performed for $\kappa \ll 1$, and at large values of $|\kappa|$ one has large contributions of irrelavant terms to $v_s$ (and also to $K$).     

{\it{TLL parameter $K$.} }
The TLL parameter $K$ was shown to be an efficient probe to capture quantum critical points in low-dimensional quantum systems~\cite{Nishimoto,Bipartite,Bipartite2}. It can be related to the magnetization fluctuation of a subsystem $A$ with length $l$, 
${\cal{F}}_L(l)=\langle(\sum_i S^z_i-\sum_i \bar{S}^z_i)^2\rangle$, where $i$ belongs to the subsystem $A$ with the average magnetization $\sum_i \bar{S}^z_i$, and the fluctuation behaves as~\cite{Song2010}
\begin{eqnarray}
 {\cal{F}}_L(l)=\frac{K}{\pi^2} \ln\left[\frac{L}{\pi}\sin\left(\frac{\pi l}{L}\right)\right]-\frac{(-1)^lb_0}{\left[\frac{L}{\pi}\sin\left(\frac{\pi l}{L}\right)\right]^{2K }}+b_1,
\label{Fluctuation}
\end{eqnarray}
where $b_0$ and $b_1$ are non-universal constants. From the bipartite fluctuations ${\cal{F}}_L(l)$ one obtains the following expression for $K$~\cite{Nishimoto}:
\begin{eqnarray}
 K(L)=\frac{\pi^2\left({\cal{F}}_L\left(\frac{L}{2}-2\right)-{\cal{F}}_L\left(\frac{L}{2}\right)\right)}{\ln\left[\cos\left(\frac{2\pi}{L}\right)\right]}.
\label{K_L}
\end{eqnarray}
In the derivation of Eq.(\ref{K_L}) we took into account the fact that for $\kappa>0$  the ${\cal O}(L^{-2K})$ correction given by the second term in Eq.(\ref{Fluctuation}) oscillates on alternating sites. Thus, ${\cal{F}}_L(\frac{L}{2})$ and ${\cal{F}}(\frac{L}{2}-2)$ is a more relevant choice. The reason of using this formula is to obtain an accurate estimnation of the TLL parameter $K$ within the parameter space $-1<\kappa \lesssim 0.33$. 
  The results of our calculations for the TLL parameter $K$ are presented in Fig.~\ref{fig:VandK} (red squares). Remarkably, the analytical result of bosonization (solid line) and DMRG results (red squares for $L=128$, $|K(L=128)-K(L=96)|\sim 10^{-4}$) are in agreement in the parameter range $|\kappa|<0.25$. For larger values of $|\kappa|$ the discrepancy between the two is large and grows with $\kappa$, which arises due to irrelevant terms excluded from our bosonization analysis.

\begin{figure}[t]
\includegraphics[width=\columnwidth]{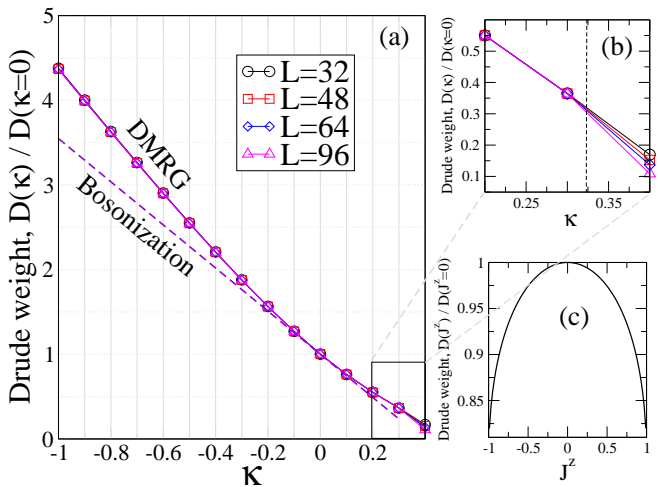}
\caption{ (a) DMRG (symbols) and bosonization (dashed line) results for the Drude weight ${\cal D}$ as functions of $\kappa$ at $-1<\kappa<-0.33$ and (b) the same in the vicinity of the critical $\kappa_c$ for various system sizes. The dashed vertical line in (b) denotes the critical $\kappa_c$. The Drude weight for the 1D XXZ as a function of the Ising interaction amplitude $J^z$ demonstrated in (c). }
\label{fig:Drude1}
\end{figure}
\section{Numerical results: the Drude weight}
\label{Sec6}
The results presented in the previous section agree to a large extent with the bosonization results for $v_s$ and $K$. In this section we show our numerical results for the Drude weight obtained for different regimes. We first demonstrate the results for the Drude weight as a function of $\kappa$.
Although one can obtain ${\cal {D}}=\frac{v_sK}{\pi}$ directly from the results of the previous calculations of $v_s$ and $K$, to affirm the results of the calculations we estimated the Drude weight ${\cal D}$ using the finite-difference Eq.(\ref{Fdiff}). The results are in agreement with ${\cal D}=\frac{v_s K}{\pi}$ and are shown in Fig.~\ref{fig:Drude1} (a) for the system sizes $L=\lbrace 32,48,64,96 \rbrace$ (symbols). The dashed line represents the bosonization result. In accordance with the bosonization result, which predicts the linear change with $\kappa$ (with the slope $-\frac{8v_F}{\pi^2}$) the Drude weight varies linearly with $\kappa$ at $|\kappa|\ll 1$. At large values of $\kappa$ the discrepancy becomes larger and the Drude weight is larger than the values predicted from the bosonization. It is important to note that $|{\cal D}(L=96)-{\cal D}(L=64)|\sim 10^{-4}$ for $-1<\kappa<0.3$, hence we accept ${{\cal D}(L=96)}$ as the thermodynamic value of ${\cal D}$. As shown in Fig.~\ref{fig:Drude1}(b), only in the vicinity of the transition point $\kappa_c=0.3256(2)$ (vertical dashed line) the Drude weight decreases with the system size noticeably. In the thermodynamic limit ${\cal D}=0$ for $\kappa>\kappa_c$, since one enters the gapped insulating phase, and the Drude weight should vanish within this phase.
\begin{figure}[t]
\includegraphics[width=\columnwidth]{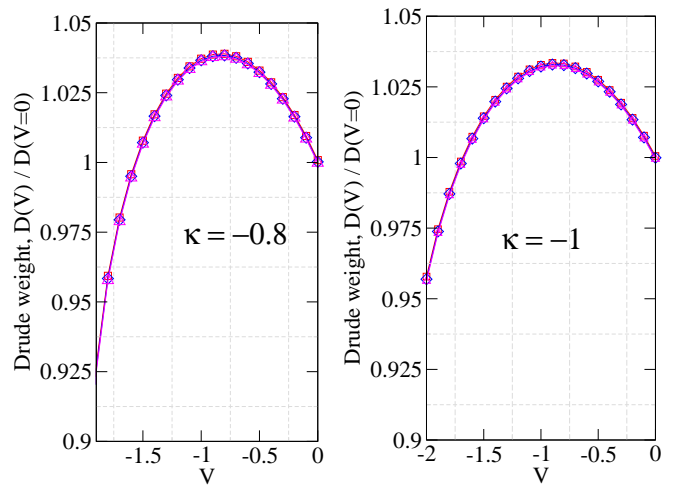}
\caption{ DMRG results for the Drude weight $\cal D$ as a function of the nearest-neighbor attraction amplitude $V$ at fixed $\kappa=-0.8$ (left panel) and $\kappa=-1$ (right panel) for various system sizes $L=\lbrace 32,48,64,96 \rbrace$. }
\label{fig:VRho}
\end{figure}

\begin{figure}[t]
\includegraphics[width=\columnwidth]{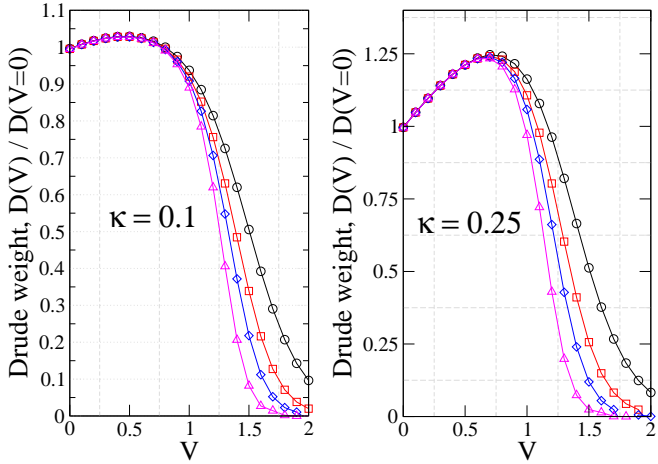}
\caption{ DMRG results for the normalized Drude weight $\cal D$ as a function of the nearest-neighbor repulsion amplitude $V$ at fixed $\kappa=0.1$ (left panel) and $\kappa=0.25$ (right panel) for various system sizes $L=\lbrace 32,48,64,96 \rbrace$. }
\label{fig:PRho}
\end{figure}

We now demonstrate interesting effects which can not be captured within the bosonization analysis and are only identified numerically. As it was mentioned in the previous sections, density-density interactions do not lead to strong renormalization of the Drude weight. If one considers $\kappa=0$ and introduces the nearest-neighbor density-density interactions $V\hat{n}_i\hat{n}_{i+1}$, then one obtains the fermionic model dual to the 1D XXZ model discussed in the previous sections. The Drude weight in this case can be calculated exactly and decreases from the non-interacting value ${\cal D}=\frac{v_F}{\pi}$ for both attractive and repulsive regimes, as shown in Fig.\ref{fig:Drude1}(c). At finite $\kappa$ an interesting effect is numerically observed. The Drude weight at a given $\kappa>0$ ($\kappa<0$) can be further enhanced, when the density-density interaction is repulsive (attractive). It first increases with the amplitude of interactions $|V|$ taking a maximum value, and then it decreases, since large $|V|$ causes phase transition to the gapped phase. The results are given in Figs.~\ref{fig:VRho}-\ref{fig:PRho}.
  
\begin{figure}[t]
\includegraphics[width=\columnwidth]{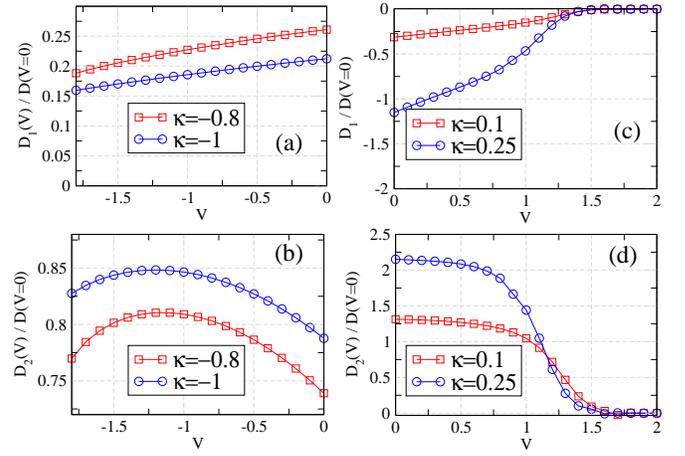}
\caption{DMRG results at fixed $L=96$ for the decomposed Drude weights (${\cal D}_1$ and ${\cal D}_2$) for the different regimes. The results for $\kappa<0$ and attractive density-density interactions are given on the left panel (a)-(b). For the other regime the plots are given on the right panel (c)-(d)}.
\label{fig:NV}
\end{figure}

If one considers negative $\kappa$, where the pair-hopping events are energetically favoured, attractive density-density interactions further enhance the Drude weight, as shown in Fig.~(\ref{fig:VRho}) at $\kappa=-0.8$ and $\kappa=-1$. In both cases the maximum ${\cal{D}}$ are reached at $V/J_1\sim -1$. Interestingly, the larger value of $|\kappa|$ is, the stronger is the effect, although in total the effect causes only $\sim 3 \%$ increase of the Drude weight. The effect in this regime can be explained as follows. For $\kappa<0$ the pair-hopping events are dominating, which results in a larger mobility of particles and in increase of the Drude weight. Single-particle hoppings, in turn, occasionally  destroy hopping pairs. Thus, if one bounds pairs by the attractive density-density interactions, pairs hop more steadily and hence the Drude weight increases. On the other hand, strong attractive interactions reduce the mobility of the particles (the excitation velocity  $v_s$ decreases with $|V|$) eventually causing phase separation at large $V/J_1\sim -2$, where $v_s=0$. To support this qualitative picture we decomposed the Drude weight ${\cal{D}}(V)={\cal D}_1(V)+{\cal D}_2(V)$, where the first term corresponds to the Drude weight obtained from the single-particle current $I_1$, whereas the second term corresponds to the Drude weight obtained from the pair current $I_2$:
\begin{equation}
    {\cal{D}}_i=-\frac{dI_i}{d\Phi}, 
\end{equation}
where $i\in\lbrace1,2\rbrace$. The corresponding plots are presented in Fig.~\ref{fig:NV} (a)-(b) (left panel). The single-particle component of the Drude weight ${\cal D}_1(V)$ decreases with the amplitude of the attractive interaction, whereas the second term exhibits a pronounced peak. These two plots demonstrate that it is only the pair current which is responsible for the non-monotonic behaviour observed in Fig.~\ref{fig:VRho}. This also supports our qualitative arguments discussed above.

The same effect is observed for positive $\kappa>0$, as shown in Fig.~\ref{fig:PRho} at $\kappa=0.1$ and $\kappa=0.25$. In the regime with $\kappa>0$ (frustrated regime), the single-particle fermionic hopping along the first leg of the ladder pins the electronic density on the other leg resulting in the decrease of the total particle mobility. Weak repulsive density-density interactions weaken localization effects, which results in a slight increase of the Drude weight, as shown in Fig.~\ref{fig:PRho}. As in the previous case, large repulsive interactions result in the phase transition to the charge density wave insulating phase. Both positive $\kappa>0$ and $V>0$ reduce the TLL parameter $K$ until it becomes $K=1/2$ and a gap opens in the single-particle spectrum.     
    Alternative explanation can be given by decomposing the total Drude weight, as in the previous regime. The results are demonstrated in Fig.~\ref{fig:NV}(c)-(d). The persistent currents along the legs of the zig-zag ladder and along the diagonal have opposite directions. This results in the the opposite signs of the decomposed Drude weights ${\cal{D}}_1$ and ${\cal{D}}_2$. The Drude weight ${\cal{D}}_1$ (with negative sign) linearly decreases with the amplitude of moderate repulsive interactions ($V<0.75$) between the neighboring sites for both values of $\kappa>0$, as expected. On the contrary, the Drude weight ${\cal D}_2$ weakly depends on the amplitude of the moderate repulsive interactions ($V<0.75$). As a result, one has pronounced peak values in Fig.~\ref{fig:PRho}.

\section{Conclusions}
To summarize, by means of the bosonization and DMRG methods, we studied effects of the interplay between the pair hoppings and density-density interactions on the Drude weight in a 1D system of spinless fermions. We first considered the effects of pair hopping on the Drude weight and found that it changes monotonically when the pair hopping amplitude is varied. In the regime of strong pair hopping the Drude weight can be strongly enhanced (almost by an order of magnitude), and hence the induced persistent currents can have large amplitudes. We also numerically demonstrated that density-density interactions can further enhance the Drude weight in the presence of the pair hopping term. This is in sharp contrast to the interaction effect by itself, which is always restricted to a reduction of the Drude weight. These findings can be tested using the platforms suggested in the recent proposals~\cite{Proposal1,Proposal2}.    
\label{Conc}
 
\begin{acknowledgements}
 We thank Michele Filippone for useful comments on the manuscript.
 This research was supported in part through computational resources of the HPC facilities at HSE University~\cite{Kostenetskiy_2021}. MSB thanks Basic Research Program of HSE for the provided support. 

\end{acknowledgements}

\begin{widetext}
\begin{appendix}

\section{Field- theoretic bosonization}
In order to obtain an effective bosonic field theory of a given fermionic model we consider the following continuous limit:

\begin{eqnarray}
    c_j \longrightarrow \sqrt{a}~ \psi(x),~~~~~x=j \cdot a,
\end{eqnarray}
where the lattice spacing $a \rightarrow 0$ and the Brillouin zone is $\left( -\frac{\pi}{a},\frac{\pi}{a} \right]\rightarrow (-\infty,\infty) $. The fermionic field operator is related to the long-wavelength part of the original fermions.

In one dimension the Fermi sphere is reduced to Fermi points. In the low energy limit we can linearize the spectrum around the Fermi points and define the right and left movers
\begin{equation}
    \epsilon_{R(L)}(k)=\pm v_F (k \mp k_F)+E_F.
\end{equation}

In order to get the low-energy effective model we expand the fermionic fields around the Fermi points
\begin{eqnarray}
\psi(x) \approx e^{i k_F x} \psi_{R }(x)+e^{-i k_F x} \psi_{L}(x),    
\end{eqnarray}

where the right and left fields are given by

\begin{equation}
  \psi_{R,L }(x) \equiv \frac{1}{\sqrt{L}} \sum_k e^{\pm i k x} c^{R,L}_k,~~~~~~c^{R,L}_k\equiv c_{\pm(k+k_F)}
  \end{equation}

For each species of  fermions we define the bosonic operators:
\begin{equation}
 b^{R,L}_q \equiv \sqrt{\frac{2 \pi }{ L q}} \rho_{R,L}(-q) ~,~~~~~b^{R,L\dagger}_{q} \equiv \sqrt{\frac{2 \pi }{L q}} \rho_{R,L}(q),~~~~~q>0,   
\end{equation}
 which obey the usual bosonic commutation relations. We define the  bosonic field operators by

 \begin{equation}
\varphi_{R,L}(x)=\pm\frac{i}{\sqrt{2 L}}\sum_{q>0} \frac{e^{\pm i q x}}{\sqrt{q}} e^{-\alpha q/2} b_q^{R,L},~~~\phi_{R,L}(x)=\varphi_{R,L}(x)+ \varphi^{\dagger}_{R,L}(x),
\end{equation}
where $\alpha$ is a large momentum cutoff. In the limit of $\alpha \rightarrow 0$ and $L \gg 1$ (which we will imply below) boson fields obey the following commutation relations:
\begin{equation}
    [\phi_{R,L}(x), \phi_{R,L}(y) ]=\pm \frac{i}{4} sign(x-y) 
\end{equation}

One can show now that the density of particles can be rewritten in terms of boson field as follows:
\begin{equation}\label{denbos}
\rho_{R,L}(x)=:\psi^{\dagger}_{R,L}(x) \psi_{R,L}(x):~=-\frac{1}{\sqrt{\pi}}\partial_x \phi_{R,L}(x).
\end{equation}

The bosonization identity reads as,
\begin{eqnarray}
    \psi_{R,L}(x)=\frac{\hat{F}_{R,L}}{\sqrt{L}} :e^{\mp i \sqrt{4 \pi} \phi_{R,L}(x)}:~=\frac{\hat{F}_{R,L}}{\sqrt{2 \pi \alpha}}  e^{\mp i \sqrt{4 \pi} \phi_{R,L}(x)},
\end{eqnarray}
where the Klein factors $\hat{F}_{R,L}$ guarantee  anticommutation relations for fermions and they commute with the bosonic operators.   

Let's define the bosonic field, which is expressed as a sum of the left and right moving  fields, and its dual field as follows:
\begin{equation}
    \phi(x)=\phi_R(x)+\phi_L(x)~~,~~~~\theta(x)=\phi_R(x)-\phi_L(x).
\end{equation}
The field $\partial_x \theta(x)$ is a canonical momentum field conjugate to $\phi(x)$
\begin{equation}
[\phi(x),\partial_y \theta(y)]=i \delta(x-y).
\end{equation}

{\it The free fermionic} part of the Hamiltonian is bosonized as follows:
\begin{flalign}
H_0=-\frac{J_1}{2}\sum\limits_{j=1}^{N}\left( c_j^\dagger c_{j+1}+c^\dagger_{j+1}c_j\right)
\longrightarrow-\frac{J_1}{2} \int dx ~\Bigl[  \psi^{\dagger}_R(x)  \psi_R(x+a) e^{i k_F a}+\psi^{\dagger}_L(x) \psi_L(x+a) e^{-i k_F a}+ \text{h.c.}\Bigr],
\end{flalign}
where in the last equation we neglect oscillating terms. Using the bosonization identity we get the free boson Hamiltonian:
 \begin{equation} 
 H_0=\frac{v_0}{2} \int dx ~\left[  :(\partial_x \theta)^2 +(\partial_x \phi) ^2: \right],
 \end{equation}
 where $v_0=J_1\cdot a$.

{\it In the presence of density-density interactions}, one-dimensional critical fermionic theories are mapped by the bosonization procedure onto a Luttinger liquid Hamiltonian:
\begin{equation}\label{Luttinger}
H= \int dx \frac{v_s}{2} \left[ K (\Pi(x))^2+\frac{1}{K} (\partial_x\phi(x))^2 \right],
\end{equation}where $\Pi(x)=\partial_x \theta(x)$ is a conjugate momentum.

Notice that $v_s$ has the dimension of velocity and is a renormalized Fermi velocity of the interacting system. The parameter $K$ is dimensionless Luttinger parameter: $K = 1$ corresponds to free fermions, whereas $K > 1$ encodes attractive fermions and $0 < K < 1$ repulsive fermions.

\section{Bosonization of $J_1-J_2$ model}

The pair hopping term in the Hamiltonian of the zig-zag model is given by:

\begin{equation}
    H_{int}=-J_2
\sum_n c_{n+1}^{\dagger} :c^{\dagger}_n c_n: c_{n-1}+h.c. \rightarrow -J_2 a \int dx~ \psi^{\dagger}(x+a):\psi^{\dagger}(x) \psi(x):\psi(x-a)+h.c.
\end{equation}
Decomposing the fermionic fields to the right and left moving fields we can rewrite the Hamiltonian in the following way:

\begin{eqnarray}\label{Hzig}
H_{int}=-J_2 a \int  dx \left[ \sum_{s =R,L}  \psi_s^{\dagger}(x+a)  \sum_{\nu=R,L}:\psi_\nu^{\dagger}(x)\psi_\nu(x): \psi_s(x-a) e^{-2i k_F a \cdot s}\right.-\nonumber \\ 
\left.-\sum_{s=R,L} \psi_s^{\dagger}(x+a)\psi_s (x)\psi_{-s}^{\dagger}(x)\psi_{-s} (x-a)+\sum_{s=R,L}\psi_s^{\dagger}(x+a)\psi_{-s}(x-a) \psi_{s}^{\dagger}(x)\psi_{-s}(x)\right] +h.c.  
\end{eqnarray}

Exploiting the bosonization identity and neglecting oscillating terms, one obtains \cite{Lecheminant2001}:

\begin{equation}
    H_{int}=-\frac{ 4~a J_2}{\pi} \int dx :\left(\partial_x \theta \right)^2: -\int dx \frac{ a J_2 }{(\pi \alpha)^2} \cos{\left(\sqrt{16 \pi} \phi \right) }
\end{equation}

Thus, the full Hamiltonian of the $J_1-J_2$ model reads
\begin{equation} 
H=\frac{v_0}{2}\int dx~\left[ :(\partial_x \phi)^2:+\left(1-\frac{8 J_2}{\pi J_1} \right):(\partial_x \theta)^2: \right] - \frac{2 g}{(2 \pi \alpha)^2 } \int dx \cos{\left(\sqrt{16 \pi} \phi \right) }.
\label{B4}
\end{equation}
Comparing Eq.~\ref{B4} with the Luttinger liquid Hamiltonian (\ref{Luttinger}), we obtain:

\begin{equation}
K=\sqrt{1-\frac{8 \kappa }{\pi}}~~~~u=v_0 \sqrt{1- \frac{8 \kappa}{\pi}}
\end{equation}
The last term in Eq. (\ref{Lutzig}) is the Umklapp scattering term with $g=2 a J_2$.   

\end{appendix}
\end{widetext}


\begin{thebibliography}{99}
\bibitem{Buttiker}
M. Büttiker, Y. Imry, and R. Landauer, \href{http://dx.doi.org/10.1016/0375-9601(83)90011-7}{Physics Letters A {\bf 96}, 365 (1983).}

\bibitem{Cheung}
H.-F. Cheung, E. K. Riedel, and Y. Gefen,\href{https://doi.org/10.1103/PhysRevLett.62.587}{ Phys. Rev. Lett. 62, 587 (1989).} 

\bibitem{Levy}
L. P. Lévy, G. Dolan, J. Dunsmuir, and H. Bouchiat, \href{https://doi.org/10.1103/PhysRevLett.64.2074}{Phys. Rev. Lett. {\bf 64}, 2074 (1990).}

\bibitem{Shanks}
A. C. Bleszynski-Jayich, W. E. Shanks, B. Peaudecerf, E. Ginossar, F. V. Oppen, L. Glazman, and J. G. E. Harris, \href{https://doi.org/10.1126/science.1178139}{Science {\bf 326}, 272 (2009).}
 
 \bibitem{Schmid}
 A. Schmid, \href{https://doi.org/10.1103/PhysRevLett.66.80}{Phys. Rev. Lett. {\bf 66}, 80 (1991).}
\bibitem{Kulik}
I. O. Kulik, \href{https://doi.org/10.1063/1.3514415}{Low Temp. Phys. {\bf 36}, 841 (2010).}

\bibitem{Viefers}
S. Viefers, P. Koskinen, P. Singha Deo, and M. Manninen, \href{https://doi.org/10.1016/j.physe.2003.08.076}{Phys. E: Low-dimensional Syst. Nanostruct. {\bf 21}, 1 (2004).}

\bibitem{Bouchiat}
H. Bouchiat, \href{https://doi.org/10.1103/Physics.1.7}{Physics 1, 7 (2008).}

\bibitem{Imry}
H. Bary-Soroker, O. Entin-Wohlman, and Y. Imry, \href{https://doi.org/10.1103/PhysRevLett.101.057001}{Phys. Rev. Lett. {\bf 101}, 057001 (2008).}

\bibitem{Shastry}
B. S. Shastry and B. Sutherland, \href{https://doi.org/10.1103/PhysRevLett.65.243}{Phys. Rev. Lett. {\bf 65}, 243 (1990).}

\bibitem{Fye}
R. M. Fye, M. J. Martins, D. J. Scalapino, J. Wagner, and W. Hanke, \href{https://doi.org/10.1103/PhysRevB.44.6909}{Phys. Rev. B {\bf{44}}, 6909 (1991).}
\bibitem{Scalpino}
D. J. Scalapino, S. R. White, and S. C. Zhang, \href{https://doi.org/10.1103/PhysRevLett.68.2830}{Phys. Rev. Lett. {\bf 68}, 2830 (1992).}

\bibitem{GiamShast}
T. Giamarchi and B. S. Shastry, \href{}{Phys. Rev. B {\bf 51}, 10915 (1995).}
\bibitem{Kohn}
W. Kohn, \href{https://doi.org/10.1103/PhysRev.133.A171}{Phys. Rev. {\bf 133}, A171 (1964).}

\bibitem{Millis}
Giamarchi and A. J. Millis, \href{}{Phys. Rev. B {\bf 46}, 9325 (1992).}

\bibitem{Sauer}
 J. A. Sauer, M. D. Barrett, and M. S. Chapman, \href{https://doi.org/10.1103/PhysRevLett.87.270401}{Phys. Rev. Lett. {\bf 87}, 270401 (2001).}

\bibitem{Gupta}
 S. Gupta, K. W. Murch, K. L. Moore, T. P. Purdy, and D. M. Stamper-Kurn, \href{https://doi.org/10.1103/PhysRevLett.95.143201}{Phys. Rev. Lett. {\bf 95}, 143201 (2005).}

\bibitem{Ryu}
C. Ryu, M. F. Andersen, P. Cladé, V. Natarajan, K. Helmerson, and W. D. Phillips, \href{https://doi.org/10.1103/PhysRevLett.99.260401}{Phys. Rev. Lett. {\bf 99}, 260401 (2007).}

\bibitem{Lesanovski}
I. Lesanovsky and W. von Klitzing, \href{https://doi.org/10.1103/PhysRevLett.99.083001}{Phys. Rev. Lett. {\bf 99}, 083001 (2007).}

\bibitem{Eckel}
S. Eckel, F. Jendrzejewski, A. Kumar, C. J. Lobb, and G. K. Campbell, \href{https://doi.org/10.1103/PhysRevX.4.031052}{Phys. Rev. X { \bf 4}, 031052 (2014).}
 
 \bibitem{Pichler}M. Łacki, H. Pichler, A. Sterdyniak, A. Lyras, V. E. Lembessis, O. Al-Dossary, J. C. Budich, and P. Zoller, \href{https://doi.org/10.1103/PhysRevA.93.013604}{Phys. Rev. A {\bf 93}, 013604 (2016).}

\bibitem{Amico}
L. Amico, A. Osterloh, and F. Cataliotti, \href{https://doi.org/10.1103/PhysRevLett.95.063201}{ Phys. Rev. Lett. {\bf 95}, 063201 (2005).}

\bibitem{Rossini}
M. Cominotti, D. Rossini, M. Rizzi, F. Hekking, and A. Minguzzi, \href{https://doi.org/10.1103/PhysRevLett.113.025301}{Phys. Rev. Lett. 113, 025301 (2014).}

\bibitem{Gallemi}
A. Gallemí, M. Guilleumas, M. Richard, and A. Minguzzi, \href{https://doi.org/10.1103/PhysRevB.98.104502}{Phys. Rev. B 98, 104502 (2018).}

\bibitem{Mancini}
M. Mancini, G. Pagano, G. Cappellini, L. Livi, M. Rider, J. Catani, C. Sias, P. Zoller, M. Inguscio, M. Dalmonte, and L. Fallani, \href{https://doi.org/10.1126/science.aaa8736}{Science 349, 1510 (2015).}

\bibitem{Genkina}
D. Genkina, L. M. Aycock, H.-I. Lu, M. Lu, A. M. Pineiro, and I. B. Spielman, \href{https://doi.org/10.1088/1367-2630/ab165b}{New J. Phys. 21, 053021 (2019).}


\bibitem{Meden}
V. Meden and U. Schollwöck, \href{https://doi.org/10.1103/PhysRevB.67.035106}{Phys. Rev. B {\bf 67}, 035106 (2003).}

\bibitem{Dias}
F. C. Dias, I. R. Pimentel, and M. Henkel, \href{https://doi.org/10.1103/PhysRevB.73.075109}{Phys. Rev. B {\bf 73}, 075109 (2006).}

\bibitem{Bouzerar}
G. Bouzerar, D. Poilblanc, and G. Montambaux, \href{https://doi.org/10.1103/PhysRevB.49.8258}{Phys. Rev. B {\bf 49}, 8258 (1994).}
 

\bibitem{Berkovitz}
R. Berkovits, \href{https://doi.org/10.1103/PhysRevB.48.14381}{Phys. Rev. B {\bf 48}, 14381 (1993).}

\bibitem{Haller}
A. Haller, M. Rizzi, and M. Filippone, \href{https://doi.org/10.1103/PhysRevResearch.2.023058}{Phys. Rev. Research {\bf 2}, 023058 (2020).}

\bibitem{Bischoff}
M. Bischoff, J. Jünemann, M. Polini, and M. Rizzi, \href{}{Phys. Rev. B {\bf 96}, 241112(R) (2017).}

  
\bibitem{RuhmanAltman}
J. Ruhman and E. Altman,
\href{https://doi.org/10.1103/PhysRevB.96.085133}{Phys. Rev. B {\bf 96}, 085133 (2017).} 
\bibitem{Mazza}
L. Gotta, L. Mazza, P. Simon, and G. Roux,
\href{https://doi.org/10.1103/PhysRevLett.126.206805}{Phys. Rev. Lett.  {\bf 126}, 206805 (2021).}
\bibitem{Mazza2}
L. Gotta, L. Mazza, P. Simon, and G. Roux,
\href{https://doi.org/10.1103/PhysRevB.104.094521}{Phys. Rev. B  {\bf 104}, 094521 (2021).}

\bibitem{White1}
S. R. White,
\href{https://doi.org/10.1103/PhysRevLett.69.2863}{Phys. Rev. Lett. {\bf 69}, 2863 (1992)}.

\bibitem{White2}
S. R. White,
\href{\doibase  10.1103/PhysRevB.48.10345}{Phys. Rev. B {\bf 48}, 10345 (1993)}.
\bibitem{Schollwock}
U. Schollwöck,
\href{https://doi.org/10.1016/j.aop.2010.09.012}{Annals of physics {\bf 326}, 96 (2011)}.
\bibitem{Rizzi}
 P. Silvi, F. Tschirsich, M. Gerster, J. Junemann,
D. Jaschke, M. Rizzi, and S. Montangero, 
\href{\doibase  10.21468/SciPostPhysLectNotes.8}{SciPost Phys.
Lect. Notes, {\bf 8}, (2019)}.

\bibitem{JWT}
	P. Jordan and E. P. Wigner, \href{\doibase 10.1007/BF01331938}{Z. Phys. {\bf 47}, 631 (1928)}.
\bibitem{Proposal1}
	M. Dalmonte, S. I. Mirzaei, P. R. Muppalla, D. Marcos, P. Zoller, and G. Kirchmair, \href{\doibase  10.1103/PhysRevB.92.174507}{Phys. Rev. B {\bf 92}, 174507 (2015)}.
\bibitem{Proposal2}
    L Barbiero, J Cabedo, M Lewenstein, L Tarruell, and A Celi
{arXiv:2212.06112 (2023).}
 
\bibitem{Kurlov2021}
 	D.V. Kurlov, S. Malikis, and V. Gritsev, \href{https://doi.org/10.1103/PhysRevB.105.104302}{Phys. Rev. B {\bf 105}, 104302 (2022)}.
 
\bibitem{Bahovadinov2022}   
M. S. Bahovadinov, D. V. Kurlov, S. I. Matveenko, B. L. Altshuler, and G. V. Shlyapnikov, \href{\doibase 10.1103/PhysRevB.106.075107}{Phys. Rev. B {\bf 106}, 075107 (2022)}.
 

\bibitem{Kurlov2023}
 P. Orlov, A. Tiutiakina, R. Sharipov, E. Petrova, V. Gritsev, and D. V. Kurlov, \href{https://doi.org/10.1103/PhysRevB.107.184312}{Phys. Rev. B {\bf 107}, 184312 (2023).}
 
 \bibitem{Motrunich2023}
 F. M. Surace and O. Motrunich,
 {arxiv:2302.12804 (2023). }

 
\bibitem{Bahovadinov2023}
M. S. Bahovadinov, R. O. Sharipov, B. L. Altshuler, and G. V. Shlyapnikov, \href{https://arxiv.org/abs/2308.10063}{arXiv:2308.10063 (2023).}

 

\bibitem{Haldane1982}
F. D. M. Haldane,
\href{\doibase 10.1103/PhysRevB.25.4925}{Phys. Rev. B {\bf 25}, 4925(R) (1982)}.

 25, 4925(R) 
 \bibitem{Okamoto1993}
	K. Nomura and K. Okamoto, \href{\doibase 10.1143/JPSJ.62.1123}{J. Phys. Soc. Jpn. {\bf 62}, 1123 (1993)}.
\bibitem{Sota2010}
	T. Sugimoto, S. Sota, and T. Tohyama, \href{\doibase 10.1103/PhysRevB.82.035437}{Phys. Rev. B {\bf 82}, 035437 (2010)}.
\bibitem{Hirata1999}
S. Hirata, and K. Nomura,
\href{\doibase 10.1103/PhysRevB.61.9453}{Phys. Rev. B {\bf 61}, 9453 (1999)}.
 \bibitem{Lecheminant2001}
P. Lecheminant, T. Jolicoeur, and P. Azaria,
\href{\doibase 10.1103/PhysRevB.63.174426}{Phys. Rev. B {\bf 63}, 174426 (2001)}.
\bibitem{Mishra2013}
T.Mishra, R. Pai, S. Mukerjee, and A. Paramekanti,
\href{\doibase 10.1103/PhysRevB.87.174504}{Phys. Rev. B {\bf 87}, 174504 (2013)}.
  \bibitem{GiamarchiBook}
 T. Giamarchi, Quantum physics in one dimension (Oxford University Press, Oxford, 2004).
\bibitem{GogolinBook}
A.O. Gogolin, A.A. Nersesyan, and  A.M. Tsvelik, Bosonization and strongly correlated systems (Cambridge University Press, Cambridge, 2004).
 \bibitem{Maslov}
D. L. Maslov, Fundamental aspects of electron correlations and quantum transport in one-dimensional systems. {arXiv:2203.0506035. (2004)}.
\bibitem{YangYang66}
C. Yang, and C. Yang,  \href{https://link.aps.org/doi/10.1103/PhysRev.150.321}{Phys. Rev. {\bf 150}, 321 (1966).} 
\bibitem{Xavier}
J. C. Xavier, \href{https://doi.org/10.1103/PhysRevB.81.224404}{Phys. Rev. B {\bf 81}, 224404 (2010).}
\bibitem{Taddia}
M. Dalmonte, E. Ercolessi, and L. Taddia, \href{https://doi.org/10.1103/PhysRevB.85.165112}{Phys. Rev. B {\bf 85}, 165112 (2012).}

\bibitem{Nishimoto}
S.~Nishimoto,
\href{\doibase 10.1103/PhysRevB.84.195108}{Phys. Rev. B {\bf 84}, 195108 (2013)}.
\bibitem{Bipartite}
H. F. Song, S. Rachel, C. Flindt, I. Klich, N. Laflorencie, and K. Le Hur,
\href{\doibase 10.1103/PhysRevB.85.035409}{Phys. Rev. B {\bf 85}, 035409 (2012)}. 
\bibitem{Bipartite2}
S. Rachel, N. Laflorencie, H. F. Song, and K. Le Hur,
\href{\doibase 10.1103/PhysRevB.108.116401}{Phys. Rev. Lett. {\bf 108}, 116401 (2012)}. 
\bibitem{Song2010}
H.F. Song, S. Rachel, and K. Le Hur,
\href{\doibase 10.1103/PhysRevB.82.012405}{Phys. Rev. B {\bf 82}, 012405 (2010)}.

 \bibitem{Kostenetskiy_2021}
 P. S. Kostenetskiy, R. A. Chulkevich  and V. I. Kozyrev, \href{https://doi.org/10.1088/1742-6596/1740/1/012050}{J. Phys.: Conf. Ser. {\bf 1740}, 012050 (2021)}.
 



\end{thebibliography}
\end{document}